\documentclass[11pt,preprint]{aastex}
\tighten 
\begin{document}
\newcommand{\lya}{Lyman~$\alpha$}
\newcommand{\lyb}{Lyman~$\beta$}
\newcommand{\za}{$z_{\rm abs}$}
\newcommand{\ze}{$z_{\rm em}$}
\newcommand{\cmtwo}{cm$^{-2}$}
\newcommand{\nhi}{$N$(H$^0$)}
\newcommand{\degpoint}{\mbox{$^\circ\mskip-7.0mu.\,$}}
\newcommand{\kms}{\,km~s$^{-1}$}      
\newcommand{\minpoint}{\mbox{$'\mskip-4.7mu.\mskip0.8mu$}}
\newcommand{\peryr}{\mbox{$\>\rm yr^{-1}$}}
\newcommand{\secpoint}{\mbox{$''\mskip-7.6mu.\,$}}
\newcommand{\sqdeg}{\mbox{${\rm deg}^2$}}
\newcommand{\squig}{\sim\!\!}
\newcommand{\subsun}{\mbox{$_{\twelvesy\odot}$}}
\newcommand{\et}{{\rm et al.}~}

\def\ltsima{$\; \buildrel < \over \sim \;$}
\def\simlt{\lower.5ex\hbox{\ltsima}}
\def\gtsima{$\; \buildrel > \over \sim \;$}
\def\simgt{\lower.5ex\hbox{\gtsima}}
\def\arcs{$''~$}
\def\arcm{$'~$}
\def\erf{\mathop{\rm erf}}
\def\erfc{\mathop{\rm erfc}}
\title{THE POPULATION OF FAINT OPTICALLY-SELECTED AGN AT $z \sim 3$ \altaffilmark{1}}
\author{\sc Charles C. Steidel\altaffilmark{2}, Matthew P. Hunt, 
and Alice E. Shapley}
\affil{California Institute of Technology, MS 105--24, Pasadena, CA 91125}
\author{\sc Kurt L. Adelberger\altaffilmark{3}}
\affil{Harvard-Smithsonian Center for Astrophysics, 60 Garden St., Cambridge, MA 02139}
\author{\sc Max Pettini}
\affil{Institute of Astronomy, Madingley Road, Cambridge UK}
\author{\sc Mark Dickinson and Mauro Giavalisco}
\affil{Space Telescope Science Institute, 3700 San Martin Drive, Baltimore, MD 21218}

\altaffiltext{1}{Based, in part, on data obtained at the 
W.M. Keck Observatory, which 
is operated as a scientific partnership among the California Institute of Technology, the
University of California, and NASA, and was made possible by the generous financial
support of the W.M. Keck Foundation.
} 
\altaffiltext{2}{Packard Fellow}
\altaffiltext{3}{Harvard Society Junior Fellow}
\begin{abstract}
We discuss a sample of 29 AGN (16 narrow-lined and 13 broad-lined)
discovered in a spectroscopic survey of $\sim 1000$ star-forming Lyman-break galaxies (LBGs) at $z \simeq 3$. 
Reaching apparent magnitudes of ${\cal R}_{\rm AB}=25.5$, the sample includes broad-lined
AGN approximately
100 times less UV luminous than most surveys to date covering similar redshifts, and the
first statistical sample of UV/optically selected narrow-lined AGN at high redshift.  
The fraction of objects in our survey with clear evidence for AGN activity is $\simeq 3$\%. A substantial
fraction, perhaps even most, of these objects would not have been detected in even the deepest existing X-ray
surveys. We argue that these AGN are plausibly hosted by the equivalent of LBGs.
The UV luminosities of the broad-lined AGN in the sample are compatible with
Eddington-limited accretion onto black holes that satisfy the locally determined
$M_{BH}$ versus $M_{\rm bulge}$ relation given estimates of the stellar masses of LBGs.
The clustering properties of the AGN are compatible with their being hosted by
objects similar to LBGs.  
The implied lifetime of the active AGN phase in LBGs, if it occurs some time during
the active star-formation phase, is $\sim 10^7$ years. 

\end{abstract}

\section{INTRODUCTION}

Until very recently, virtually all surveys for QSOs and AGN at high redshift
have been geared toward the detection of relatively ``extreme'' and
rare objects, ranging from UV excess or color selection of QSOs (e.g.,
Fan et al. 2001 ; Boyle \et 2000; Warren, Hewett, \& Osmer 1994) 
to the selection of objects by their extreme radio 
power (e.g., White et al. 2000; Dunlop \& Peacock 1990, McCarthy 1993). 
These surveys have been quite successful because they make
use of selection techniques that can be applied over large swaths of
sky, and for which spectroscopic follow-up has proved straightforward
and highly efficient due to the optical brightness of the sample. 
There are few constraints on broad-lined AGN drawn from the faint end of the
UV/optical luminosity distribution (the faintest published survey is still
that of Koo \& Kron 1988, reaching the equivalent of ${\cal R} \sim 21$ at
$z \sim 3$); these fainter objects are expected to dominate the
production of photons that, together with emission from young galaxies,  maintain the
ionization of the intergalactic medium at intermediate redshifts ($z \sim 1-5$). 
Thanks to results from the Chandra X-Ray Observatory (Mushotzky \et 2000;
Barger \et 2001; Alexander \et 2001), we are just beginning
to learn about the high redshift ($z > 2$) population of narrow-lined, 
obscured AGN of ``type II''.  These objects 
can be quite unobtrusive at the optical (rest-UV) or radio wavelengths
used by most AGN surveys in the past. Despite their possible importance to the
overall AGN demographics, only a handful has been identified to date  
(e.g., Stern \et 2002a, Norman \et 2002). 

At the same time, a very strong case has been developed in the local universe
for a tight correlation between the properties of the stellar populations
in bulges and spheroids and the mass of central black holes (e.g., 
Kormendy \& Richstone 1995, Magorrian \et 1998, Merritt \& Ferrarese 2001). 
The presence of this correlation over a very wide range of mass scale
strongly suggests that the formation of the spheroid stellar populations
and the central black hole are causally linked. One might then reasonably
expect that the era during which the spheroid stars were formed might
also be that during which the black holes were most likely to be accreting
material from gas-rich environs. The ``quasar era'' is now known to be
rather strongly peaked near $z \sim 2.5$, declining rapidly 
at both higher and lower redshifts (e.g., Boyle \et 2000; Warren, Hewett, \& Osmer 1994; 
Schmidt, Schneider, \& Gunn 1995;
Kennefick, Djorgovski, \& de Carvalho 1995; Fan \et 2001; Shaver \et 1999). 
Since it is now possible to routinely observe star-forming galaxies near the 
peak of the ``quasar era'', there is the opportunity to assess the level
of AGN activity that is ongoing as the galaxies are undergoing rapid star formation.

In this paper, we present some initial results on the AGN component
of a moderately large spectroscopic survey for galaxies selected by their large unobscured 
star formation rates at redshifts $z\sim 3$. The survey should contain all
but the most heavily obscured star forming galaxies at such redshifts (Adelberger
\& Steidel 2000), and
as we detail below, is well-suited for detecting active
accretion power in the same star forming objects because of the selection
criteria used and the large number of spectra obtained. For the first time
at high redshift, it may be possible to assess the fraction of rapidly
star forming galaxies that are simultaneously playing host to significant
accretion power.  In any case, the survey has uncovered a relatively large
number of UV-selected, relatively faint, broad-lined and narrow-lined AGN, 
whose properties we summarize.  

We assume a cosmology with $\Omega_m=0.3$, $\Omega_{\Lambda}=0.7$, and
$h=0.7$ throughout. 

\section{THE AGN SAMPLE}

\subsection{ General Properties and Sample Definitions}

The full LBG survey consists of 16 independent fields covering a total of 
0.38 square degrees. The effective volume covered by 
a color-selected survey depends in a relatively complex way on the color,
magnitude, and redshift of objects in the targeted sample 
(Steidel \et 1999; Adelberger 2002). 
We have presented some of the details of the LBG
selection function in Steidel \et 1999; the AGN described in this paper 
satisfied precisely the same photometric criteria as the star forming galaxies in 
the sample.
The complete details of the LBG photometric and
spectroscopic survey will be presented in Steidel \et 2002 (in preparation); here we
concentrate on the small sub-sample for which there is spectroscopic evidence
for the presence of AGN.

At redshifts $z \sim 3$, the distinctive colors of Lyman break objects 
depend largely on properties of the intervening intergalactic medium (IGM)
where the mean free path of photons shortward of 912 \AA\ in the rest
frame is short, resulting in a pronounced drop in flux in the observed $U_n$ band
even for objects whose spectra do not have intrinsic breaks at the Lyman
limit (e.g., Madau 1995, Steidel \& Hamilton 1993). 
Because objects in the spectroscopic sample were selected without regard
to morphology (i.e., no attempt was made to remove point sources from the
catalog), we expect that our spectroscopic sample should be at least as
complete for objects dominated by non-stellar emission as compared
to normal star forming galaxies.  

The complete catalog of LBG candidates in the $z \sim 3$ survey fields consists
of 2440 objects in the apparent magnitude range $19.0 \le {\cal R} \le 25.5$.
Simulations suggest that only $\sim 50$\% of objects with LBG-like intrinsic colors
at $z \simeq 3 \pm 0.3$ will be included in our color-selected photometric
catalogs due to various sources of incompleteness (Steidel \et 1999).
While we attempted to obtain uniform data in each field, variations
in Galactic extinction, seeing, sky brightness, and integration time would require each field to
be treated independently in a proper evaluation of completeness. 
Over the full survey, we spectroscopically observed a total of 1344 objects (55\%
of the photometric sample), of which 51 are identified as Galactic stars, 988 
are high redshift objects with $<z>=2.99\pm0.29$, and 306 remain unidentified. 
We classified an object as a broad-lined AGN
if its spectrum contained any emission line with FWHM$>2000$ \kms. Such objects
always contained several broad emission lines, 
usually at least Lyman $\alpha+$NV, Si~IV $\lambda 1397$,
C~IV $\lambda 1549$, and C~III] $\lambda 1909$. 
Of the 13 broad-lined AGN identified, two show evidence for ``associated'' broad
absorption lines of high ionization species such as O~VI, N~V, Si~IV, and C~IV. 
Objects were classified as narrow-lined AGN if their strong Lyman $\alpha$ emission
was accompanied by significant emission in C~IV $\lambda 1549$ and if no
emission line had FWHM $>2000$ \kms. These objects usually had He~II $\lambda 1640$ and
C~III] $\lambda 1909$, and often detectable N~V $\lambda 1340$ and O~VI $\lambda 1034$.  
Given the quality of the typical LBG survey spectrum,
the narrow C~IV emission line would have to exceed equivalent widths of a few \AA\ in
the rest frame, or line fluxes of $\sim 2 \times 10^{-17}$ ergs s$^{-1}$ cm$^{-2}$,
to have been recognized.  

The coarse properties of the two AGN samples are summarized in Table 1.  

The composite spectra of both classes of AGN, formed by shifting each spectrum into
the rest frame based on the emission line redshift, 
normalizing by the median continuum level in the rest-frame 1600-1800 \AA\ range,
and averaging, are shown in figure 1. The top panel of figure 1 also shows for comparison the composite 
spectrum of much brighter
QSOs (typically $V \sim 18$, from Sargent, Steidel, \& Boksenberg 1989 and Stengler-Larrea \et 1995) 
that would satisfy the same LBG color selection criteria.
We note the striking similarity of the two QSO samples, which are separated by
a factor of $\sim 100$ in UV luminosity. There is clear evidence for the
Baldwin (1977) effect in the increasing strength of the C~IV emission line with
decreasing continuum luminosity, and the faint QSO composite has a much more
prominent narrow He~II $\lambda 1640$ emission line. 

Our definition of a narrow-lined AGN, based on the detected presence
of high ionization emission lines, is admittedly somewhat arbitrary; however, it  
relies on the fact that the vast majority of LBGs show no detectable emission
in C~IV and He~II even when Lyman $\alpha$ emission is strong, and that it is
difficult to produce significant nebular lines of these high ionization species without
the hard ionizing spectrum of an AGN component. 
The line ratios observed among the objects identified as
narrow-lined AGN are quite similar to those of local Seyfert 2 galaxies--in fact,
for the composite spectrum shown in the bottom panel of figure 1, the Lyman $\alpha$/C~IV
and Lyman $\alpha$/C~III] ratio are essentially identical to a composite Seyfert 2
spectrum presented by Ferland \& Osterbrock (1986). The composite narrow-lined
AGN spectrum is also strikingly similar, in both continuum and emission line
properties, to the composite spectrum of high
redshift radio galaxies (from Stern \et 1999)

A separate, but related, issue is how the narrow-lined, UV--selected AGN fit in with 
the so-called ``type II'' QSOs discovered
recently in deep Chandra images. The optical spectra of faint X-ray sources
detected with Chandra are quite diverse, ranging from broad-lined AGN to objects
without clear UV/optical indicators of the presence of AGN (e.g., Barger \et 2001,
Hornschemeier \et 2001, Mushotzky 2002 and references therein). However, for the
relatively small number of published spectra of identified Chandra sources at $z > 2$,
every one would have satisfied either our ``broad-lined'' or ``narrow-lined''
AGN criteria.  In particular, the spectra of the 2 published ``type II'' QSOs at
high redshift (Stern \et 2002a, Norman \et 2002) both resemble
our faint optically selected narrow-lined AGN spectra in their UV emission line
properties. As discussed below, at present 
there is only limited information on the X-ray emission from objects in the optically selected samples. 
Clearly, optical/UV selection of AGN will impose different selection criteria than
X-ray selection, and there may well be strong X-ray emitting AGN at $z >2$ that 
would either not be detected at all in the rest-UV, or not be recognizable as AGN from 
their UV spectra, because of heavy obscuration. Similarly, because of greater absolute
sensitivity in the UV and widely varying UV/X-ray flux ratios (for whatever reason), 
it may be that faint AGN selection in the UV can identify objects whose X-ray fluxes
are still beyond the deepest Chandra integrations. We discuss this issue further in \S 3.

\subsection{Estimates of Internal Completeness of the AGN Sub-Sample}

While a more detailed analysis of the AGN selection function
and a derivation of the UV luminosity function of faint AGN are deferred to
another paper (Hunt \et 2002, in preparation), here we make some approximate statements on 
the completeness of our spectroscopic AGN sample. 
When we could not determine a redshift from a spectrum we had obtained, it
was usually due to an absence of
emission lines and inadequate continuum signal to measure the relatively
weak absorption lines which help establish redshifts for a large fraction of
our galaxy sample. Since every AGN in our sample has several strong emission
lines (including strong Lyman $\alpha$ emission), it
is unlikely that an AGN with UV-detectable features 
in the target redshift range 
would not have been recognized even in spectra of much lower than average quality.
We can estimate an upper limit on the number of unrecognized narrow-lined AGN (for reasons of inadequate
S/N in the spectra) among those objects identified as normal star-forming galaxies by
examining high S/N composite spectra of the LBGs (with identified AGN excluded).  
The average intensity of C~IV emission in the
narrow-lined AGN sample is 
$\sim 20\%$ that of Lyman $\alpha$ (see figure 1). If we assume that this ratio is
characteristic of narrow-lined AGN that we failed to flag as such,  and we use
the fact that the intensity ratio of Lyman $\alpha$ to C~IV emission in the spectral
composite of non-AGN LBGs is $\simgt 100$ for the quartile of the LBG sample having the strongest 
Lyman $\alpha$ emission strength\footnote{This is the only sub-sample of the LBGs
that has a mean Lyman $\alpha$ equivalent width close to that of the AGN}, then an upper 
limit on the fraction of AGN-like spectra to have contributed
to that sub-sample is $\sim 5/100=5$\%. 
The corresponding limit on the fractional contribution of unrecognized 
AGN to the full LBG sample would then be $\sim 0.25*5\simeq 1$\%.
The true fraction with unrecognized AGN-like spectra is likely to be smaller than this limit, since in most individual spectra 
we could have recognized C~IV emission at the level seen in the composite AGN spectrum, and
because low-level C~IV emission (part of which is due to the stellar P-cygni feature) 
is expected in the rapidly star forming galaxies
even without AGN excitation.  

Thus, we expect that, with respect to the photometric LBG sample, the spectroscopic
AGN sample is close to $N_{\rm obs,spec}/N_{\rm phot}= 1344/2440\sim 55$\% complete, i.e., that any AGN in the
LBG photometric sample that was
attempted spectroscopically would have yielded a redshift. 
We estimate that our present spectroscopic AGN sample contains only $\sim 30$\% of the AGN
in our fields with $2.7 \simlt  z \simlt 3.3$ and satisfying our photometric criteria
${\cal R} \le 25.5$, $G-{\cal R} < 1.2$, and $U_n-G > (G-{\cal R})+1 $
(i.e., the spectroscopic completeness of 0.55 times the estimated photometric completeness of $\sim 50$\%).
The mean redshift of the narrow-lined AGN in our sample is somewhat different from that
of the galaxies, probably due to a combination of the subtleties of how the emission lines have
affected the broad--band photometry and the redder continuum color (see below). 
The broad-lined AGN completeness within the 
photometric sample is expected to be smaller than that of galaxies  
at a given redshift and apparent magnitude because, at $z \sim 3$, only about 60\% of (bright) 
QSO spectra have
intervening optically thick Lyman limit systems at high enough redshift to produce the distinctive UV color
that we depend on to identify them (see Sargent \et 1989). Galaxies do not suffer
this form of incompleteness because they are expected to have significant intrinsic
Lyman limits from a combination of stellar spectral energy distributions and opacity
to their own Lyman continuum radiation from the interstellar medium\footnote{We cannot rule out the
possibility that faint broad-lined AGN are subject to increased internal Lyman continuum
opacity compared to the bright QSOs that have been studied to date; however,
the absence of detectable interstellar absorption lines in the composite broad-lined
QSO spectrum does suggest that the typical H~I column density within the host galaxies along
our line of sight is significantly smaller than that for typical LBGs.}.  
Again, a careful treatment of these effects is deferred
to Hunt \et 2002, but this additional source of incompleteness for broad-lined AGN 
is likely to be of roughly the same order as the 
spectroscopic advantage which AGN enjoy when they are
observed. 

With these {\it caveats} in mind, a reasonable estimate of the fraction of AGN among objects in our LBG 
sample is approximately the same as the fraction of AGN within the
spectroscopically confirmed sample: $29/988$, or $\sim 3$\%. This number would increase
to $\sim 4$\% allowing for the maximal incompleteness of the narrow-lined sample discussed above. The
observed ratio of narrow-lined to broad-lined AGN, $N(NL)/N(BL)=1.2 \pm 0.4$, is consistent  
with the ratio of broad-lined and narrow-lined radio-loud AGN found
by Willott \et 2000, although we emphasize that our numbers are not yet corrected
for relative incompleteness.   

\section{X-Ray Properties of Optically Faint AGN at $z \sim 3$}

At present, there is only a small amount of information on the 
X-ray properties of the optically faint AGN
in our sample. A cross-correlation of our LBG survey with the 1 Msec
exposure of the Chandra Deep Field North (the HDF North region) yields X-ray detections for
4 of the 148 candidates{\footnote{None of these objects is detected in the radio continuum with
the VLA (Nandra \et 2002)} ($\sim 3\%$) in an $8.7 \times 8.7$ arc minutes field centered on the deep HST
pointing (Nandra \et 2002). Of these, two have not yet been observed spectroscopically. 
The other two are a
faint broad-lined AGN with ${\cal R}=24.15$ and $z=3.406$ (HDF-oC34$=$J123633.4$+$621418), 
and a narrow-lined AGN
with ${\cal R}=24.84$ and $z=2.643$ (HDF-MMD12$=$J123719.9$+$620955). Both of these AGN have 
rest-frame 2--10 keV luminosities (uncorrected for intrinsic absorption) of $\sim 5 \times 10^{43}$
ergs s$^{-1}$. Our spectroscopic Lyman break
galaxy sample also contains a clear narrow-lined AGN spectrum, shown in figure 2, that is undetected in
the deep Chandra image and thus has an unobscured X-ray luminosity of $\simlt 5\times 10^{42}$
ergs s$^{-1}$ in the 2-10 keV band. Thus, while we expect that a large fraction of the optically
faint AGN in our sample would be detected in the deepest Chandra exposures, there is likely
also to be a sub-sample that is relatively X-ray faint that would not be detected in even the
deepest X-ray pointings to date.   Given the overall completeness estimate of $\sim 30$\%
discussed above, we expect $\sim 20$
optically faint (${\cal R} \sim 21-25.5$) 
AGN (of which $\sim 11$ would be broad-lined objects) over a redshift interval of $\Delta z \simeq
0.6$ near $z =3$ per 17\arcm\ by 17\arcm\ Chandra ACIS field
\footnote{The number 20 is just $29/(0.3*0.38) \simeq 250$ AGN per square degree. We note, however,
that the region in which 4 LBGs were directly detected with Chandra was only $\sim 25$\% of the
full Chandra field of view.}. 
Ostensibly this number is
significantly larger than the number of AGN (in a similar
redshift range) identified with
Chandra sources in the deep fields (e.g., Barger \et 2001, Hornschemeier et al 2001, Stern et al 2002b,
Crawford \et 2002), although the numbers are small
and the follow-up of optically faint Chandra sources is still underway. 
More complete surveys of both Chandra sources and faint optically--selected AGN in the
same fields will significantly improve our understanding of the overall
AGN demographics at high redshift.

\section{Are LBGs the hosts of the optically faint AGN?}

It would be interesting in the context of understanding the history and efficiency
of accretion-powered luminosity in galaxies if one could 
verify that the AGN in the sample are
hosted by the equivalent of LBGs.\footnote{Analysis of HST images of the host
galaxies of a small sample of $z \sim 2$ radio quiet QSOs by Ridgway \et (2001)
indicates that the hosts are consistent with the luminosities and morphologies
of LBGs.} 
The range of continuum apparent magnitudes (i.e., the lack of objects
brighter than ${\cal R} \sim 23$) of the narrow-lined AGN in
the spectroscopic sample is similar to that of the non-AGN LBGs in the
sample (table 1).  
The strength of
the few interstellar absorption lines that are not strongly masked by emission lines
in the composite narrow-lined AGN spectrum are quite similar to those of a composite
spectrum of non-AGN LBGs with Lyman $\alpha$ seen in emission (see figure 3).
Unfortunately, the strongest stellar feature in the spectra of LBGs is the C~IV
P-Cygni profile, which is badly affected by C~IV emission in the composite 
narrow-lined AGN spectrum. We cannot say with certainty whether the continuum light
of the narrow-lined AGN is produced by stellar or non-stellar emission-- making this
distinction is notoriously difficult even for nearby Seyfert galaxies (e.g., Gonzalez-Delgado \et 1998). 
However, the far-UV continuum slope ($\beta = -0.4$, where $f_{\lambda} \propto \lambda^{\beta}$) 
of the composite narrow-lined AGN is redder 
than all but $\sim 10$\% of the LBG sample, and is much redder than the subsample
of LBGs with similarly strong Lyman $\alpha$ emission, as illustrated in figure 3.  
If the continuum were attributed to starlight in the same manner as other LBGs then
the implied extinction would be a factor of $\sim 50$ (Adelberger \& Steidel 2000).  
At this time, only one of the narrow-lined AGN has been observed in the K band, HDF-oMD49
(see figure 2). This object has ${\cal R}-K=4.20$, making it the third-reddest LBG in the observed
sample of 118 (Shapley \et 2001).  

The brightest broad-lined object in the sample is  
more than 2 magnitudes brighter than the brightest
narrow-lined object, and there are 5 broad-lined AGN brighter than the
brightest narrow-lined AGN. 
There is some evidence for a flatter magnitude distribution
for the broad-lined AGN than for the narrow-lined AGN,  but
small number statistics and numerous possible selection effects prevent us from
making too much of this trend at this time. The relative absence of very UV-bright
narrow-lined AGN is at least qualitatively consistent with the possibility that much
of the AGN-produced UV continuum is obscured. 
There is considerable overlap between our survey fields and planned deep surveys
with the {\it Space Infrared Telescope Facility} (SIRTF), so that it should soon
be possible to measure many of these optically selected AGN at mid-IR wavelengths,
where any bolometrically luminous obscured AGN are expected to be quite prominent. 

Assuming that the broad-lined AGN are the less-obscured versions of similar AGN activity,
let us suppose for the sake of argument that the mass of putative LBG black holes 
scales with stellar bulge mass according to the relation established locally,
$M_{BH} \sim 2\times 10^{-3} M_{\rm bulge}$ (e.g., Ho 1999). Adopting the range of inferred stellar
masses of LBGs in our survey from Shapley \et 2001, we expect typical black hole masses
of $3 \times 10^7$ M$_{\sun}$ and a range from $2 \times 10^6$ to perhaps
$2\times 10^8$ M$_{\sun}$. If these black holes were radiating at the Eddington limit,
they would be expected to have observed\footnote{
Here we assume a radiative efficiency of $\epsilon \sim 0.1$
for the accretion and that $\nu L_{\nu}(UV)$ is a reasonable approximation to
the bolometric luminosity.}
${\cal R} \simeq 20.0 - 24.7$, close
to the observed range (for the broad-lined AGN) of ${\cal R}=20.6-24.8$. 
Apparently, LBGs would be capable of hosting broad-lined AGN with the range of
observed UV luminosities.

A more quantitative test of whether the AGN and the LBGs share the same
host dark matter halos comes from an evaluation of the
clustering statistics of the AGN with respect to LBGs. 
We have performed tests, using methods similar to those outlined
in Adelberger \et 2002,  of the density of LBGs
around AGN as compared to that of other (non-AGN) LBGs. 
Evaluated on scales 
$\delta z =0.0085$ and $\delta \theta=200$ arc minutes, we find that the density of
LBGs in the vicinity of narrow-lined AGN is $0.96\pm 0.24$ times the density
of LBGs around other LBGs. The density of LBGs around broad-lined AGN is
$1.58\pm 0.33$ times higher than the density of LBGs around other LBGs. 
These crude tests suggest that the narrow-lined AGN cluster very similarly to non-AGN LBGs, 
and that broad-lined AGN may be more strongly clustered than typical LBGs. 
A more in-depth
treatment of these issues will be presented in Hunt \et 2002.   

In any case, it seems plausible
that the observed AGN may be hosted by the equivalent of LBGs. If this is indeed the
case, the fraction of LBGs in which obvious AGN activity is present may provide a rough
timescale for near-Eddington accretion rates onto central black holes, as follows. 
The characteristic timescale for star formation episodes in LBGs is estimated
to be $\sim 300$ Myr, inferred from the modeling of the far-UV to
optical (rest-frame) colors (Shapley \et 2001; Papovich, Dickinson, \& Ferguson 2001).
If the 3\% AGN activity reflects the duty cycle of 
significant black hole accretion in LBGs, it would imply
an active accretion timescale of $\sim 10^{7}$ years, broadly consistent with
the expected black hole masses given the implied Eddington mass accretion rate
of $\sim 1$ M$_{\sun}$ yr$^{-1}$.  AGN lifetimes of this order have been inferred
from theoretical studies of black hole growth based on mergers in hierarchical
models of structure formation (e.g. Kauffmann \& Haehnelt 2000)
and from consideration of the QSO luminosity functions and the distribution of
black hole masses in the local universe (e.g., Haehnelt \et 1998, Yu \& Tremaine 2002). 
Significantly longer accretion timescales of $\sim 500$ Myr have been suggested by Barger \et (2001) 
based on the observation that $\sim 4\%$ of ``$L^{\ast}$ galaxies at all redshifts'' are   
X-ray sources, but such timescales may refer to a very different, more protracted
process of sub-Eddington accretion onto black holes in well-formed galaxies primarily
at $z < 1$. 

\section{Discussion}

While we defer more quantitative statements to a future paper, there are
several statements we can make that are unlikely to change after more
careful modeling of incompleteness. First, narrow-lined AGN that are
identifiable optically using LBG color selection criteria are
quite common, with a space density (at $z \sim 3$) $\sim 50-100$ times larger than that
of the spectroscopically--similar high redshift radio galaxies (cf. Willott \et 2001). 
The implied surface density of AGN per square degree per unit redshift at $z \sim 3$
reaches $\sim 400$.   
It is still uncertain, due to small number statistics and incomplete surveys, what fraction of this number 
would also be Chandra sources that may contribute significantly to the X-ray background\footnote{About
half of the narrow-lined AGN in our sample would be classified as ``optically faint'', defined
by Alexander \et 2001 to be objects with $I > 24$, or $I_{AB}\simgt24.5$.}. 
Nevertheless, we can say with some confidence that narrow-lined AGN make a negligible contribution
to the $z \sim 3$ UV background: the total $1500$ \AA\ luminosity of narrow-lined AGN in our
sample is only $\sim 20$\% that contributed by the broad-lined  AGN, and $\simlt 2$\%
of that contributed by non-AGN LBGs. We do not know the narrow-lined AGN contribution to the
{\it ionizing} UV background, but it is likely to be far smaller than 20\% of the broad-lined AGN 
background judging by the red continuum colors and the expectation that the objects
are fairly heavily obscured. 

Using currently accepted parameterizations of the $z \sim 3$ QSO luminosity
function (Pei 1995), $\sim 75$\% of the AGN-produced ionizing radiation field
would come from QSOs that have apparent magnitudes in the range ${\cal R}=20-25.5$,
and only $\sim 7$\% comes from brighter QSOs. Thus, while our sample of broad-lined  
AGN is fairly small, it extends far deeper than existing QSO surveys\footnote{
The faintest broad-lined AGN in our sample are considerably fainter than the traditional 
dividing line between QSOs and Seyfert 1 nuclei of $M_B = -23$. This absolute
magnitude, for the adopted cosmology, corresponds to ${\cal R} \simeq 23$.}, 
allowing for the first time
a direct measurement of the AGN contribution to the ionizing radiation field
at high redshift.  While it is possible that star formation in LBGs may dominate
the production of Lyman continuum photons at $z \sim 3$ (Steidel, Pettini, \& Adelberger
2001), broad-lined AGN such as those in our faint sample almost certainly provide 
a substantial fraction of the higher energy photons which apparently reionized He~II near $z \sim 3$.  
An accurate measurement of the  AGN ionizing photon production 
requires careful attention to issues
of photometric and spectroscopic completeness, which will be presented in
Hunt \et 2002 (in preparation). 

\bigskip
\bigskip
CCS, MPH, and AES have been supported by grant
AST-0070773 from the U.S. National Science Foundation and by the David and Lucile
Packard Foundation. KLA acknowledges support from the Harvard Society of Fellows.
We thank Daniel Stern for providing the composite radio galaxy spectrum, and
for useful comments on an early draft. Useful conversations with Paul Nandra and Aaron Barth  
are also acknowledged.

\begin{deluxetable}{lccc}
\tablewidth{0pc}
\tablecaption{Properties of the AGN Among the $z \sim 3$ LBG Sample}
\tablehead{
\colhead{} & \colhead{Broad-Lined}  & \colhead{Narrow-Lined} & \colhead{LBGs\tablenotemark{a}}
} 
\startdata
Number  & 13  & 16 & 959 \\ 
$\langle z \rangle \pm \sigma_z$  &  $3.03 \pm0.35$ & $2.67 \pm0.35$ & $2.99\pm 0.30$ \\
$\langle {\cal R} \rangle $\tablenotemark{b} & $23.3$ & $24.4$ & $24.6$ \\
${\cal R}$ Range & $20.6-24.8$  & $22.6-25.4$ & $22.8-25.5$ \\
\enddata
\tablenotetext{a}{Excluding objects identified as AGN.}
\tablenotetext{b}{Mean AB magnitude at an effective wavelength of 6830 \AA\, or
$\simeq 1700$ \AA\ in the rest frame at $z\simeq 3$.  }
\end{deluxetable}

\bigskip
\newpage
\begin{figure}
\epsscale{0.9}
\plotone{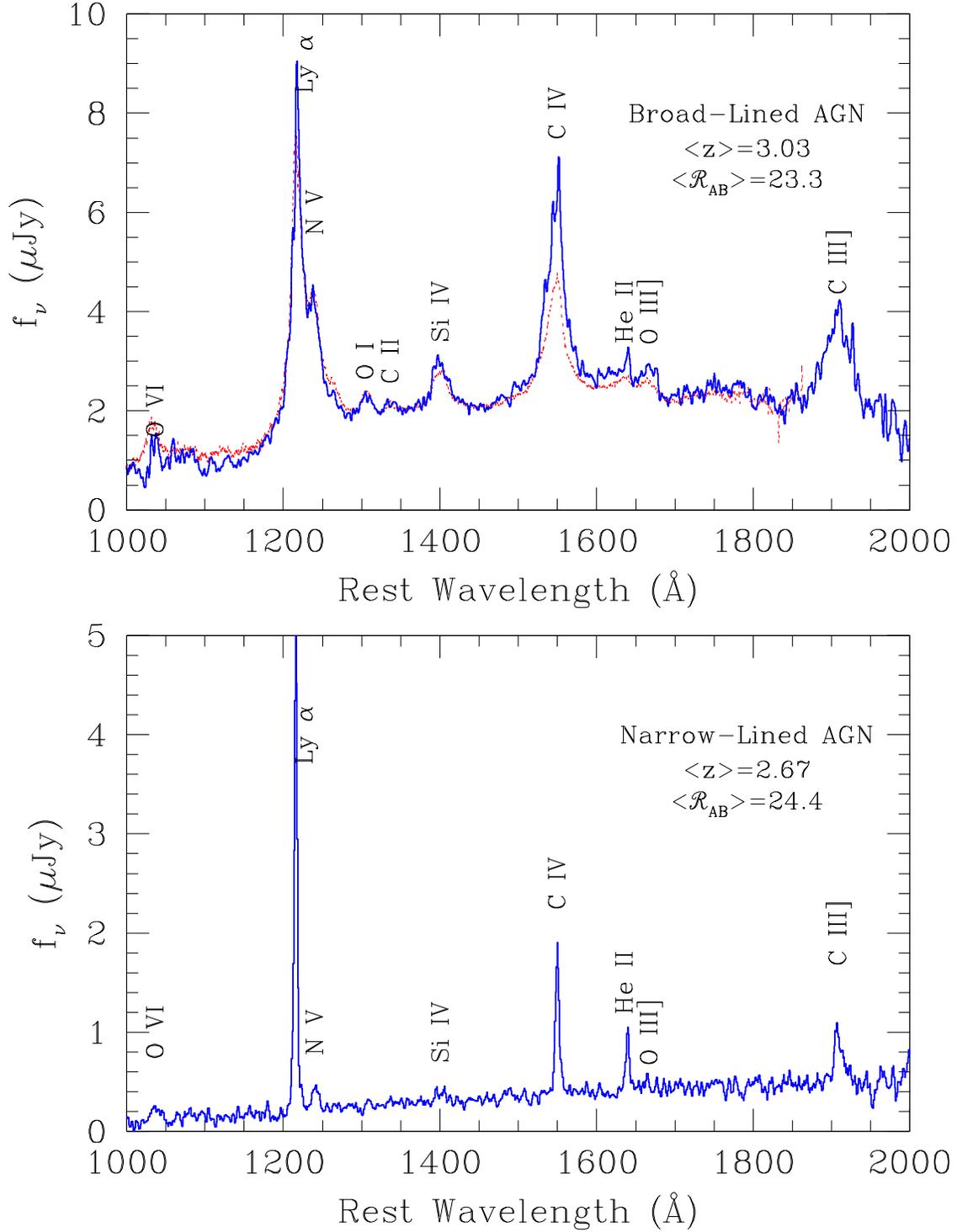}
\caption{a) The composite spectra of the 13 broad-lined AGN (blue). Superposed
in red (dotted) is the (scaled) composite spectrum formed from a sample of QSOs that is 100 times brighter on average (Sargent, Steidel, \& Boksenberg
1989; Stengler-Larrea \et 1995).  b) The composite spectrum formed
from the 16 narrow-lined AGN identified in the LBG spectroscopic sample. 
}
\end{figure}

\begin{figure}
\plotone{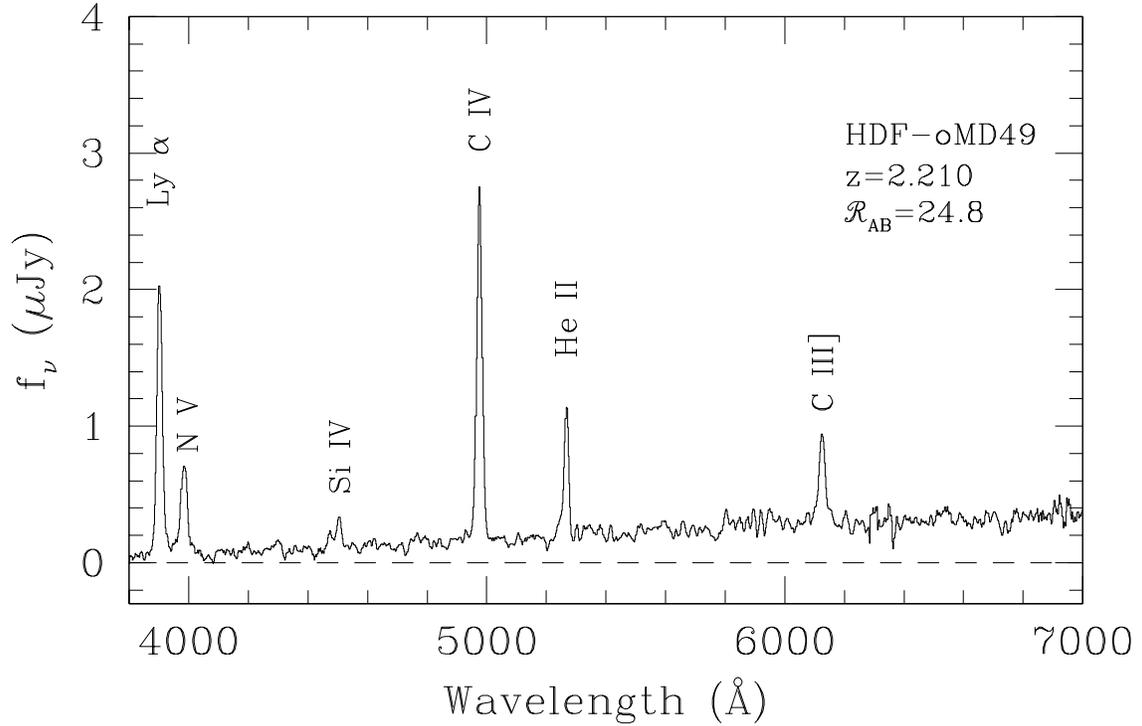}
\caption{The observed spectrum of HDF-oMD49, a narrow-lined AGN that is undetected in
the 1 Ms Chandra Deep Field North X-ray image (coordinates: (J2000) $12~37~4.34~ +62~14~46.2$). This object has an unusually weak
Lyman $\alpha$ emission line with respect to the high ionization NV and C~IV lines,
possibly indicating a large amount of extinction in the narrow line region. 
} 
\end{figure}

\begin{figure}
\plotone{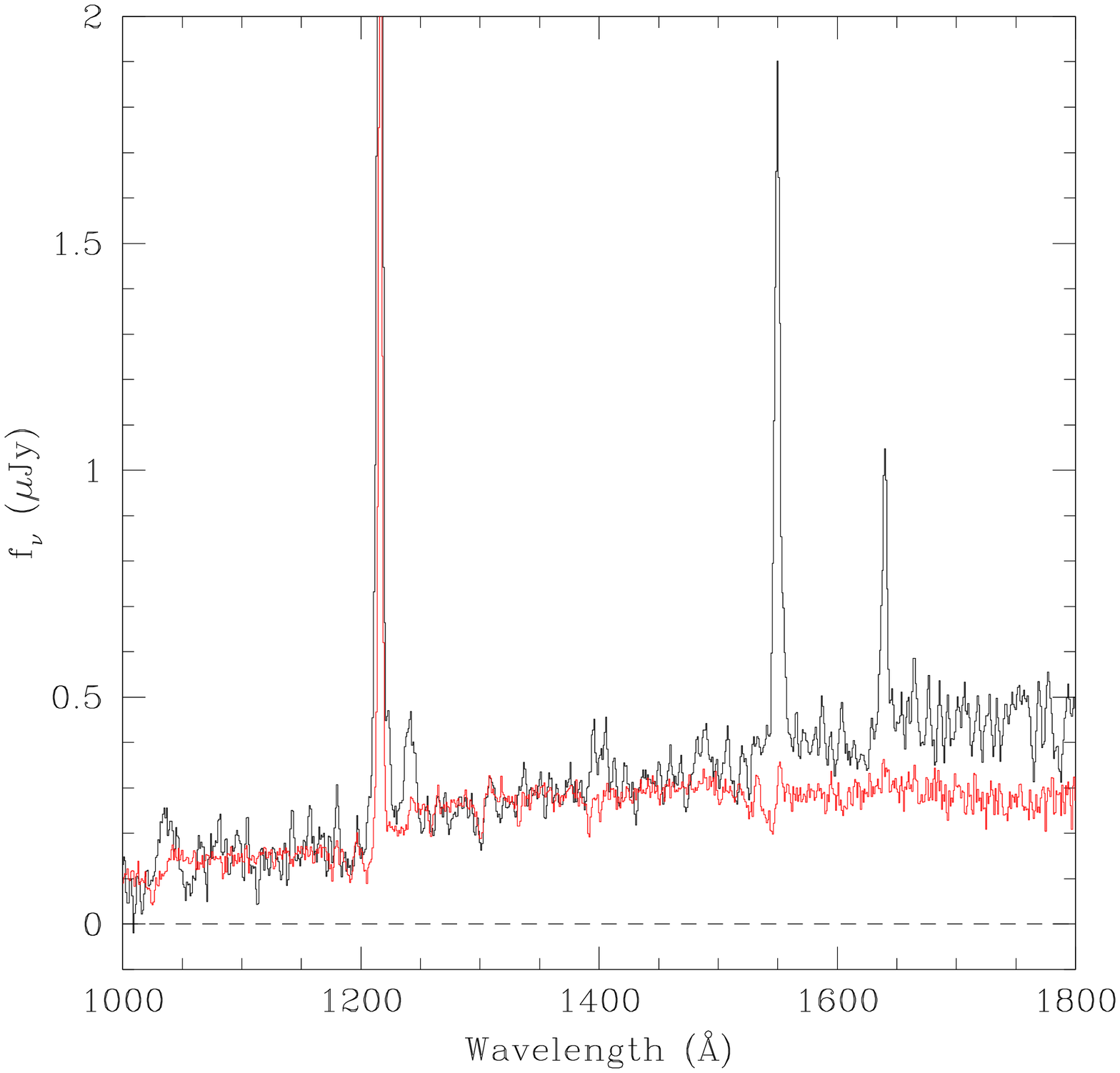}
\caption{Comparison of the continua of the composite (non-AGN) LBG spectra culled from
the subset with strong Lyman $\alpha$ emission (red; Shapley et al 2002), 
compared to that of the identified
narrow-lined AGN (black). Note that the AGN spectrum has much stronger emission lines and is
considerably redder in the continuum. Note the similarity of the strength of the
interstellar OI+SiII feature near 1303 \AA.
}
\end{figure}


\begin{references}
\reference{} Adelberger, K.L, Steidel, C.C., Shapley, A.E., \& Pettini, M. 2002, ApJ,
submitted
\reference{} Adelberger, K.L. 2002, Ph.D. thesis, California Institute of Technology
\reference{} Adelberger, K.L, \& Steidel, C.C. 2000, ApJ, 544, 218
\reference{} Alexander, D.M., Brandt, W.N., Hornschemeier, A.E., Garmire, G.P., Bauer, F.E.,
Griffiths, R.E. 2001, AJ, 122, 2156
\reference{} Baldwin, J.A. 1977, ApJ, 214, 679
\reference{} Barger, A.J., Cowie, L.L., Mushotzky, R.F., \& Richards, E.A. 2001, AJ, 121, 662
\reference{} Barger, A.J., Cowie, L.L., Bautz, M.W., Brandt, W.N., Garmire, G.P., 
Hornschemeier, A.E., Ivison, R.J., \& Owen, F.N. 2001, AJ, 122, 2177
\reference{} Boyle, B.J., Shanks, T., Croom, S.M., Smith, R.J., Miller, L., Loaring, N.,
\& Heymans, C. 2000, MNRAS, 317,1014
\reference{} Crawford, C.S., Gandhi, P., Fabian, A.C., Wilman, R.J., Johnstone, R.M., Barger, A.J.,
\& Cowie, L.L. 2002, MNRAS, in press (astro-ph/0106067). 
\reference{} Dunlop, J.S., \& Peacock, J.A. 1990, MNRAS, 247, 1990
\reference{} Ferland, G.J., \& Osterbrok, D.E. 1986, ApJ, 300, 658
\reference{} Fan, X., et al. 2001, AJ, 121, 54 
\reference{} Gonzalez-Delgado, R., Heckman, T.M., Leitherer, K., Meurer, G., Krolik, J.,
Wilson, A., Kinney, A., \& Koratkar, A. 1998, ApJ, 505, 174
\reference{} Haehnelt, Natarajan, \& Rees 1998, MNRAS, 300, 817
\reference{} Ho, L.C., in `Observational Evidence for Black Holes in the Universe', ed.
S.K. Chakrabarti (Dordrecht: Kluwer), 153
\reference{} Hornschemeier, A.E. et al 2001, ApJ, 742,777
\reference{} Kauffmann, G., \& Haehnelt, M. 2000, MNRAS, 311, 588
\reference{} Kennefick, Djorgovski, \& de Carvalho 1995, AJ, 110, 2553
\reference{} Koo, D.C., \& Kron, R.G., ApJ, 325, 92
\reference{} Kormendy, J., \& Richstone, D. 1995, ARAA, 33, 581
\reference{} Madau, P. 1995, ApJ, 441, 18
\reference{} Magorrian, J., et al. 1998, AJ, 115, 2285
\reference{} McCarthy, P.J. 1993, ARAA, 31, 639
\reference{} Merritt, D., \& Ferrarese, L. 2001, MNRAS, 320, L30
\reference{} Mushotzky, R.F. 2002, in New Views of the X-ray Universe in the XMM-Newton and
Chandra Era, in press (astro-ph/0203310)
\reference{} Mushotzky, R.F., Cowie, L.L., Barger, A.J., Arnaud, K.A. 2000, Nature, 101, 159
\reference{} Nandra, K., Mushotzky, R.F., Arnaud, K., Steidel, C.C., Adelberger, K.L., Gardner,
J.P., Teplitz, H.I., \& Windhorst, R.A. 2002, ApJ, submitted
\reference{} Norman, C.A., et al. 2002, ApJ, in press (astro-ph/0103198)
\reference{} Papovich, C., Dickinson, M., \& Ferguson, H.C. 2001, ApJ, 559, 620
\reference{} Pei, Y.C. 1995, ApJ, 438, 623
\reference{} Ridgway, S.E., Heckman, T.M., Calzetti, D., \& Lehnert, M. 2001, ApJ, 550, 122
\reference{} Sargent, W.L.W., Steidel, C.C., \& Boksenberg, A. 1989, ApJS, 69, 703
\reference{} Schmidt, M., Schneider, D.P., \& Gunn, J.E. 1995, AJ, 110, 68
\reference{} Shapley, A.E., Steidel, C.C., Adelberger, K.L., Dickinson, M., Giavalisco, M.,
\& Pettini, M. 2001, ApJ, 562, 95
\reference{} Shapley, A.E., Steidel, C.C., Adelberger, K.L., \& Pettini, M. 2002,
in 'A New Era in Cosmology', in press (astro-ph/0112262)
\reference{} Shapley, A.E., Steidel, C.C., Adelberger, K.L., Dickinson, M., Giavalisco, M., \& Pettini, M. 2001,
ApJ, 562, 95
\reference{} Shaver, P.A., Hook, I.M., Jackson, C.A., Wall, J.V., \& Kellermann, K.I.
1999,
in Highly Redshifted Radio Lines, ASP Conf Series Vol 156, eds. C.L. Carilli,
S.J.E. Radford, K.M. Menten, \& G.I. Langston, 163
\reference{} Steidel, C.C., Pettini, M., \& Adelberger, K.L. 2001, ApJ, 546, 665
\reference{} Steidel, C.C., Adelberger, K.L., Giavalisco, M., Dickinson, M., \& Pettini, M. 1999,
ApJ, 519, 1
\reference{} Steidel, C.C., \& Hamilton, D. 1993, AJ, 105, 2017
\reference{} Stengler-Larrea, E.A., et al. 1995, ApJ, 444, 64 
\reference{} Stern, D., et al. 2002a, ApJ, in press (astro-ph/0111513)
\reference{} Stern, D., et al 2002b, AJ, in press (astro-ph/0203392)
\reference{} Stern, D., Dey, A., Spinrad, H., Maxfield, L., Dickinson, M., Schlegel, D., \&
Gonzalez, R.A. 1999, AJ, 117, 1122
\reference{} Warren, S.J., Hewett, P., \& Osmer, P.S. 1994, ApJ, 421, 412
\reference{} White, R.L., et al 2000, ApJS, 126, 133
\reference{} Willott, C.J., Rawlings, S., Blundell, K.M., Lacy, M., \& Eales, S.A. 2001, MNRAS, 322, 536
\reference{} Willott, C.J., Rawlings, S., Blundell, K.M., \& Lacy, M. 2000, MNRAS, 316, 449
\reference{} Yu, Q., \& Tremaine, S. 2002, MNRAS, submitted (astro-ph/0203082).
\end{references}
\end{document}